\documentclass[fleqn,twoside]{article}
\usepackage{espcrc2}
\usepackage{epsfig}
\usepackage{amssymb}

\def\bc{\begin{center}}
\def\ec{\end{center}}
\newcommand{\be}{\begin{equation}} 
\newcommand{\ee}{\end{equation}} 
\newcommand{\bea}{\begin{eqnarray}} 
\newcommand{\eea}{\end{eqnarray}} 
\newcommand{\agg}{A_{\gamma\gamma}}

\newcommand{\pigg}{\Pi_{\gamma\gamma}}
\newcommand{\piggp}[1]{\mbox{$\Pi^{\mbox{\scriptsize{(#1)}}}_{\gamma\gamma}$}}
\newcommand{\piggpms}[1]{\mbox{$\hat\Pi^{\mbox{\small{(#1)}}}_{\gamma\gamma}$}}
\newcommand{\mt}{m_{\scriptstyle t}^2}
\newcommand{\mwo}{m_{\scriptstyle W}}
\newcommand{\mzo}{m_{\scriptstyle Z}}
\newcommand{\mw}{m_{\scriptstyle W}^2}
\newcommand{\mz}{m_{\scriptstyle Z}^2}

\newcommand{\ehs}{\mbox{$\hat{e}^{2}$}}
\newcommand{\alphah}{\mbox{$\hat{\alpha}$}}
\newcommand{\ms}{\mbox{$\overline{ MS}$}}

\title{The electric charge at the two-loop level in the Standard Model 
and the precision measurements of $\Delta\alpha_{had}$}

\author{Alessandro Vicini
\address{Dipartimento di Fisica, Universit\`a degli Studi di Milano,
  Via Celoria 16, 20133 Milano (Italy)}}

\begin{document}

\begin{abstract}
The complete calculation of the 2-loop electroweak corrections to the
renormalization of the electric charge in the Standard Model allows
to discuss in detail the value of the $\ms$ effective coupling
$\ehs(\mz)$.
We discuss the phenomenological impact of these results 
in view of the increasing accuracy in the measurements of the
cross-section of the reaction $e^+e^-\to~hadrons$ at low energies
\end{abstract}

\maketitle

\section{Introduction}
The precision tests of the Standard Model (SM) have reached in the
past years an incredibly high level of accuracy: several observables
are measured at the per mille level and two key quantities ( the $W$
boson mass $\mwo$ and the sinus of the effective weak mixing angle
$\sin^2\theta_{eff}^{lep}$ ) are even known with an error of few parts
in $10^{-4}$. The prospects for the measurements of these two
latter observables at future colliders (Tevatron run IIb, LHC, TESLA)
will further improve the present situation \cite{baur}.
In order to have a sensible comparison between the data and the
theoretical prediction, the complete calculation of the 2-loop
electroweak (EW) corrections to all relevant observables is a mandatory step.\\
The input parameters in the gauge sector of the SM lagrangian are
chosen in order to minimize the parametric error that they induce on
any theoretical prediction and are typically
the three best measured quantities $(\alpha, G_{\mu}, \mzo)$, the fine
structure constant, the Fermi constant derived from the muon decay and the mass
of the $Z$ boson; in particular $\alpha$ is known to the
impressive accuracy of 3.7 parts per billion \cite{pdg}.
On the other hand, the effective coupling which appears in the
expression of any scattering amplitude is the running electromagnetic
coupling $\alpha(s)$, but unfortunately this quantity is not so
well known as $\alpha(0)$ for two different reasons: $i)$ the presence
of non-perturbative hadronic contributions, usually indicated with
$\Delta\alpha_{had}$,  which affect the running
and $ii)$ the error due to missing higher-order perturbative corrections.\\
The precise value of $\Delta\alpha_{had}$ 
is directly related, via a dispersion relation, 
to the measurements of the total cross-section 
of the reaction $e^+e^-\to~hadrons$ \cite{procFJ}.
The latter are reaching a very high level of accuracy and make
questionable if the perturbative contributions to the running of
$\alpha$ are fully under control.\\
Recently the 2-loop renormalization of the electric charge in
the EW Standard Model has been calculated (we refer to \cite{dv} for
all technical details) and using these results the running of the
$\ms$ electric charge has been studied. We review in the present paper
the most relevant features of that calculation and discuss its
phenomenological consequences, in view of a precision measurement of
$\Delta\alpha_{had}$.
\section{The electric charge renormalization}
In pure QED the
natural definition of an effective QED coupling at the scale $\sqrt{s}$
\bea
\label{alphaqed}
\alpha (s) &=& \frac{\alpha}{1 - \Delta \alpha (s)} \\
\label{deltaal}
\Delta \alpha (s) &=&4 \pi \,\alpha {\rm Re}
\left[ \pigg (s) - \pigg(0) \right]~,
\eea
is given in terms of the photon vacuum polarization function evaluated
at different scales.\\
In the full SM, the bosonic contribution to the photon vacuum
polarization at high momentum transfer is, in general, not gauge-invariant.
Thus it cannot be included in a sensible way in Eq.(\ref{alphaqed}).
Although,  Eq.(\ref{alphaqed}) with only the fermionic contribution included
is a good effective coupling at the $\mz$ scale, it is clear that at the 
energy scales tested by the future accelerators an effective QED coupling
will have to take into account also the bosonic contributions.\\
A different definition of a QED effective coupling can be obtained
by considering the \ms\ QED coupling constant at the scale
$\mu$ defined by
\be
\alphah (\mu) = \frac{\alpha}{1 +4\pi \alpha \hat{\Pi}_{\gamma \gamma}(0)}~.
\label{alfamu}
\ee
Eq.(\ref{alfamu}) is expressed in terms of  
the on-shell electric charge counterterm which is a gauge-invariant
quantity and includes bosonic and fermionic contributions.
In the Backgroun Field Method (BFM), 
as it will be explained in the rest of the paper,
the electric charge counterterm is expressed in terms of the
$q^2 =0$ photon vacuum polarization function.\\
The electric charge renormalization has been discussed 
at the one-loop level in \cite{sir80,marsir}; at the two-loop level
it has been studied in several papers in relation to the 
$m_{\scriptscriptstyle W}-m_{\scriptscriptstyle z}$ interdependence
\cite{deltar}; in the framework of the BFM explicit results have been
presented in \cite{dv}. 

\begin{figure}[h]
\centering
\epsfig{file=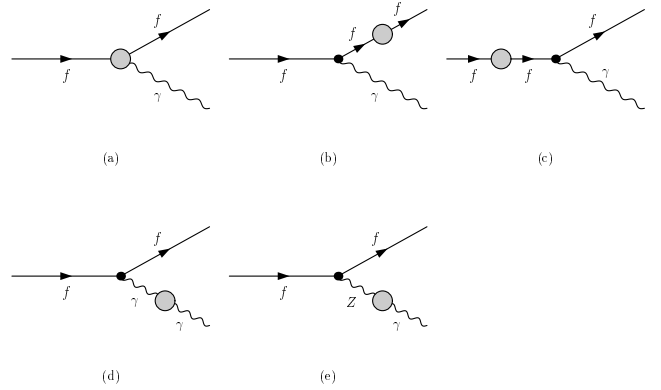,height=6cm,width=9cm,
        bbllx=150pt,bblly=300pt,bburx=600pt,bbury=600pt}
\caption{The 
diagrams of the Thomson scattering
 }
\label{fig1}
\end{figure}

The electric charge is usually defined in terms of the Thomson
scattering amplitude, i.e. the process of emission of a real photon
off a fermion, with vanishingly small energy.
The on-shell electric charge counterterm is defined as to cancel,
order by order, all divergences and all finite corrections which arise
through virtual corrections.
The classes of virtual diagrams which contribute to this amplitude are
schematically depicted in Fig.\ref{fig1}.
In QED the Ward-Takahashi identity (WI) implies a cancelation between
diagrams (a), (b) and (c), and therefore the relation between the bare
and the renormalized charge is given, via Dyson summation, by
$e^2 = e_0^2/(1-e_0^2 \piggp{f}(0))$, where $\piggp{f}(0)$ is related to
the transverse part $\agg$ of the photon self-energy via 
$\agg(q^2) = e_0^2 q^2 \pigg (q^2)$ 
and $f$ indicates the fermionic contributions.\\
The QED WI does not hold in the SM quantized in the 't Hooft-Feynman
gauge, because of a non-trivial relation between the bosonic
self-energies and the vertex corrections. Both groups of diagrams are
separately gauge dependent and only their combination yields a gauge
invariant expression, as expected for the electric charge counterterm
\cite{sir80}.
Due to the gauge dependence, the Dyson summation of the bosonic
self-energies is not manifestly evident in a renormalizable $R_{\xi}$ gauge.
At the 2-loop level, the calculation of the Thomson scattering
amplitude in the 't Hooft-Feynman gauge is difficult, due to the
presence of irreducible vertex corrections and to non-trivial
cancellations between the self-energy contributions.\\
The BFM has been introduced in \cite{BFM} 
as an alternative approach to quantize a gauge field theory.
The fields of the classical lagrangian are splitted into two component
${\mathcal L}(\hat\phi)\to {\mathcal L}(\hat\phi+\phi)$ where the symbol
$\hat\phi$ indicates all classical fields, while $\phi$ represents the
quantum fluctuations, i.e. the integration variables of the path
integral. A gauge-fixing term is introduced to break only the gauge
invariance of the quantum-fields $\phi$. The effective action remains
invariant under gauge transformations of the background fields and, as
a consequence, the Green's functions with external background fields
satisfy simple WIs.
A convenient choice of the gauge-fixing can be exploited to derive
simple QED-like WI, now in the full electroweak SM.
It has been shown \cite{ddw} that the following relations hold to all
orders in perturbation theory, for arbitrary values of the gauge parameter.
\vskip-0.5truecm
\bea
\label{QEDWard}
q^{\mu} \Gamma^{\gamma {\bar f} f}_{\mu}(q,{\bar p},p) &=& 
-e Q_f
\left[
\Sigma_f({\bar p}) - \Sigma_f(-p)
\right]
\\
\label{GGLzero}
B_{\gamma\gamma}(0) &=& 0\\
\label{GZLzero}
B_{\gamma Z}(0) &=& 0
\eea
where  $\Gamma^{\gamma {\bar f} f}_{\mu}$
is the three-point function $\gamma {\bar f} f$, 
$\Sigma_f$ is the two-point function of the fermion, $q= {\bar p} + p$
the photon momentum 
and $Q_f$ is the charge of the fermion $f$ in units $e$. 
Eq.(\ref{QEDWard}) is the usual QED-like Ward identity.
$B_{\gamma\gamma}$ and $B_{\gamma Z}$ denote the longitudinal parts of
the $\gamma\gamma$ and $\gamma Z$ self-energies. 
From the validity of eqs.(\ref{GGLzero},\ref{GZLzero}) and
from the 
analyticity properties of the two-point functions, it follows that, to all 
orders,
\vskip-0.5truecm
\bea
\label{GGTzero}
A_{\gamma\gamma}(0) &=& 0\\
\label{GZTzero}
A_{\gamma Z}(0) &=& 0~.
\eea
The validity of the WIs yields a substantial computational
simplification:
in fact, it is possible to verify that the contributions of diagrams
(a), (b) and (c) of Fig.\ref{fig1} now add up to zero and that
diagram (e) is vanishing. In the BFM, the electric charge
renormalization is due only to the photon vacuum polarization. 
From the gauge invariance implied by eq.\ref{GGTzero} also for the
bosonic contributions, it is possible to Dyson resum also these terms.\\
The relation between bare and renormalized charge is
\bea
    e^2 &=& {e^2_0 \over 1 - e_0^2 \pigg(0) }~,
\label{eq1SM} \\
\pigg(0) &=& \piggp{f}(0) + \piggp{b}(0)~,
\label{pis}\\
\piggp{f}(0) &=& \piggp{lep}(0)  + \piggp{5}(0) +  \piggp{pert}(0)~,
\label{pif}
\eea
where $\piggp{b}$ indicated the bosonic contributions.
A further distiction is in order for the fermionic part, where
$\piggp{lep}$ indicates the diagrams with one leptonic loop, 
$\piggp{5}$ those with one light quark loop exchanging a photon or a
gluon
and $\piggp{pert}$ the ones with one top loop or with one light-quark
loop, exchanging a $W$ or a $Z$ boson.
The term $\piggp{5}(0)$ can not be evaluated perturbatively, due to
strong interactions at low energy.
Rewriting $\piggp{5}(0) =  {\rm Re} \left( \piggp{5}(0)-\piggp{5}(\mz) \right)
+ {\rm Re} \piggp{5}(\mz)$, the round bracket can be evaluated via a dispersion
relation from the experimental data for the total cross-section of
$e^+e^-\to hadrons$ at low energies \cite{allworld}, 
while the second term can be
calculated perturbatively \cite{fks}, due to the large scale $q^2=\mz$.
In our 2-loop calculation we have separated the diagrams with one
light quark loop and the exchange of a photon/gluon, from those with a
light quark loop and the exchange of a $W$ or a $Z$. The latter, due
to the large scales $\mw,~\mz$, can be evaluated perturbatively at
$q^2=0$. In contrast, the former have to be treated resorting to the
experimental low energy data.\\
The complete 
analytical expressions for 
the 2-loop photon vacuum polarization are rather long and are
presented in \cite{dv}.
\section{ The $\ehs(\mz)$ $\ms$ parameter}
\begin{figure}[h]
\epsfig{file=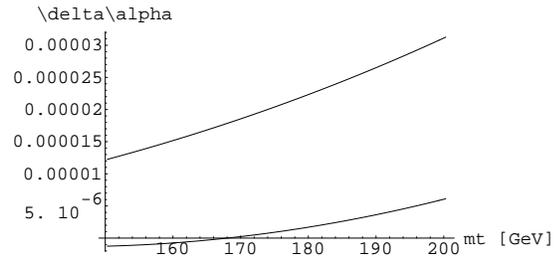,height=5cm,width=5cm,
        bbllx=00pt,bblly=0pt,bburx=200pt,bbury=200pt}
\caption{Contributions due to the top quark. The upper line shows only
  the ${\cal O}(\mt)$ terms, while the lower line is the full result}
\label{figtop}
\end{figure}

The relation given by Eq.(\ref{eq1SM}) allows to determine one
of the fundamental parameter of the \ms\ renormalization scheme,
$\ehs(m_Z)$, i.e. the \ms\ electric charge defined  at scale $m_Z$. 
The  \ms\ renormalization procedure is  defined  as the 
subtraction of pole terms of the form $(n-4)^{-m}$, where $m$ is an integer 
$\geq 1$, $n$ is the number of dimensions and the identification
of the 't~Hooft parameter $\mu$ 
with the relevant mass scale, in this case $m_Z$.\\
In order to obtain the relation between \ehs\ and $e^2$, one
writes $e_0^2 = \ehs/\hat{Z}_{e}$ in Eq.~(\ref{eq1SM}), and uses the
counterterms present in $\hat{Z}_{e}$ to cancel the $(n-4)^{-1}$ terms
in the regularized but unrenormalized vacuum polarization function
$\pigg(0)$  setting $\mu = m_Z$ in the explicit expressions \cite{fks}. 
We define $\piggpms{i}$
the self-energy expression subtracted of its divergent part $I_i/(n-4)$.
Without implementing any decoupling we have
\be
  \hat{Z}_e = 1
 +{\alphah\over4\,\pi}\left(I_l+I_t +I_5 + I_b\right){1\over n-4}
 \label{E7d}
\ee
so that
\be
  e^2 = {\ehs\over1 + (\alphah/\alpha) \Delta \alphah(\mz)} ,
\label{E8a}
\ee
with 
\bea
\label{E8b}
&&  \Delta \alphah (\mz)=  \Delta\alpha_{had}^{(5)} 
+ \frac{\alpha}\pi \left[ \frac{55}{27} \right. \\
&& \left.        + \left(\frac{11\alphah_s(\mz)}{9\pi}
 + \frac{35\alphah(\mz)}{108\pi}\right) 
   \left(\frac{55}{12} - 4 \zeta(3) \right) \right] \nonumber\\
&&-4\, \pi \alpha
\left[ \piggpms{lep}(0) + \piggpms{bos}(0) +\piggpms{pert}(0) \right] 
\nonumber
\eea
In the first line we use the value
$4 \pi^2 {\rm Re} \left(\piggp{5}(0) - \piggp{5}(\mz)\right) =
0.027690 \pm 0.000353$ \cite{jegerlehner},
for the hadronic non-perturbative contributions.
In the first and second line there is the contribution of $\piggp{5}(\mz)$
due to a loop of light quarks interacting
with an internal photon or gluon \cite{fks}.
In the last line we indicate all perturbative contributions, whose
 explicit expressions can be found in \cite{dv}.
Eq.(\ref{E8b}) can be easily solved  for \ehs, obtaining
\be
  \ehs(\mz) = {e^2\over1 - \Delta\alphah(\mz)} .
\label{E8c}
\ee
Using the following values (in GeV) for the fermion masses
$
m_e = 0.000511, m_{\mu} = 0.105658  , m_{\tau} = 1.777 , m_{t} = 174.3
$
and for the gauge bosons $m_{\scriptstyle Z} = 91.187,  
m_{\scriptstyle H} = 150$,
we have evaluated $\ehs(\mz)$.\\
\begin{table}[h]
\begin{tabular}{|c|c|c|c|c|}
\hline
              &1NP    &2 QCD  &2 QED&2 EW   \\
\hline
leptons      & 3529.2   &       &  7.66    & 10.18   \\
\hline
bosons       &   -140.7   &       &          &  -1.79   \\
\hline
top          &   -133.7   & 8.66  &  0.19   &  0.08       \\
\hline
$\piggp{5}(0)\Big|_{EW}$ &       &      &         &   4.56   \\
\hline
${\rm Re}\, \piggp{5}(\mz)$ & 473.4     & -2.39      &  -0.04  &        \\
\hline\hline
$\Delta\alpha_{had}^{(5)}(\mz)$ &  2769.0   &  \multicolumn{3}{c|}{ }        \\
\hline\hline
total        & 6497.2   &  6.27     &    7.81      &     13.03    \\
\hline
\end{tabular}
\caption{Numerical results for $\Delta\alphah(\mz)$, expressed
  in units $10^{-5}$.
The input parameters and the different groups of contributions are
  specified in the text. 
}
\label{numbersmz}
\end{table}\\
In table \ref{numbersmz}
we present separately the contributions from the leptons, from the
purely bosonic diagrams,  from the diagrams involving the top quark,
from the diagrams with light quarks exchanging a massive vector
boson (indicated with $\piggp{5}(0)\Big|_{EW}$) and from
the diagrams with light quarks exchanging a photon or a gluon,
evaluated at $q^2=\mz$. 
In the first column we consider the 1-loop and the non-perturbative
contributions.
In the other columns we distinguish the 2-loop QCD and purely QED
corrections \cite{fks}, and in the last column the full 2-loop EW
(QED+weak) corrections.
We have checked, in the lepton and in the top case, that the
appropriate subset of diagrams from our results 
reproduces the numbers presented in \cite{fks}.
Concerning the 2-loop EW diagrams involving a top quark, approximate
results including all terms of order ${\mathcal O}(\alpha^2 \mt/\mw)$
were already available \cite{dg} and could also be reproduced.\\
The largest contributions are due to light fermions (leptons and
quarks) exchanging massive vector bosons and have both positive sign.
In contrast the 2-loop purely bosonic diagrams have  negative sign
and are smaller in size. 
The size of the full 2-loop EW results is more than 13 parts in units
$10^{-5}$ and almost half of it is due to purely electroweak effects.\\
{\bf $\ms$ vs. on-shell}\\
The use of the BFM makes evident that the electric charge
renormalization is due, also in the EW SM, only to the photon vacuum
polarization.
In the on-shell scheme 
the Dyson resummation of the self-energy corrections is not possible
for the bosonic contributions at an arbitrary energy scale $q^2$,
because they would break gauge invariance.
Only at $q^2=0$ the WIs guarantee that such terms are a gauge
invariant subset.
On the other hand, the QED and QCD fermionc corrections are
a gauge invariant subset which can be
included in the definition of the on-shell running coupling
and evaluated at any energy scale.\\
The $\ms$ renormalization scheme does not suffer of this problem,
because the running of the charge is determined only by the
self-energy corrections evaluated at $q^2=0$.
The relevance of this approach becomes evident when we want to
evaluate the effective charge at high energy scales, like those of LEP2
($\sqrt{s}\geq 160$ GeV) or higher, where the massive gauge bosons are
active degrees of freedom which significantly contribute to the running.\\
{\bf Two-loop corrections and $\Delta\alpha_{had}$}\\
In the past, the error which affects the determination of
$\Delta\alpha_{had}$ has been considered a bottleneck of all precision
tests of the SM. The new results from VEPP, Da$\Phi$ne, BES anf BaBar
\cite{newexp}
are now
improving the present status. There are prospects to lower the error
on $\Delta\alpha_{had}$ from the present 
$0.027690 \pm 0.000353$ by a factor 5 to $0.00007$  or even to $0.00005$.
\cite{jegerlehner}.
``Theory-driven'' 
analyses, which use a theoretical input to improve the description of the 
cross-section $e^+e^-\to~hadrons$ at low energy,
present already today an error equal to $0.00012$ \cite{dh}.\\
We remark that the full 2-loop EW corrections shift the central value of
$\Delta\alpha$ by +0.00013 and
are of the same order of magnitude of the error which affects the
determination of $\Delta\alpha_{had}$.
The inclusion of the 2-loop EW radiative corrections for the relevant
observables is therefore a mandatory step for all precision tests of the SM.
The effect of the 2-loop EW corrections is twofold:
$i)$ they shift the central value of $\ehs(\mz)$ and 
$ii)$ reduce the theoretical perturbative uncertainty on its determination, 
which is now pushed at the 3-loop level.
In the $\ms$ scheme, 
the quantity $\ehs^2(\mz)$ thus provides a realistic estimate of the
effect of the 2-loop EW corrections which modify the running of the
effective electric charge.\\
{\bf Perturbative contributions of quarks}\\
The numerical results presented in Table \ref{numbersmz}
require two further comments.
The quasi-vanishing of the diagrams with the top quark, for realistic values of
in the input paramenters, is due to a fortuitous numerical
cancellation.
The asymptotic expansion of the top contributions in powers of
$\mt/\mw$ is not converging very rapidly, because the expansion
parameter, the top Yukawa coupling, is of the same order of magnitude
as the gauge couplings \cite{dgv}.
One verifies that the so called leading-terms ${\cal O}(\mt)$ have the
same size as the so called sub-leading ones.
Having opposite sign, they almost cancel as shown in fig.\ref{figtop}.\\
\noindent
The perturbative contributions due to light quarks and leptons,
exchanging a $W$ or a $Z$ boson, are numerically relevant: the zero
mass of the fermions compensate the suppression due to a heavy
mass. They contain a logarithmic term $\log(m_{\scriptstyle V}/\mu)~~~
(V=W,Z)$
which becomes relevant at high energy scales $\mu$.\\
{\bf The running of $\alphah$ at higher energies}\\
The running of $\alphah$ at higher energy scales can be
evaluated by changing the value of the renormalization scale $\mu$ 
in eq.\ref{E8b}.
\begin{table}[h]
\begin{tabular}{|l|c|c|c|c|}
\hline
$\mu$  & 1NP   & 2QCD & 2EW  & 
   $\hat\alpha^{-1}(\mu)$  \\
\hline
~$m_{\scriptstyle Z}$ & 6485.42 &  6.27 & 13.03 &  128.11 $\pm$ 0.05 \\
\hline
~300     & 6991.91 & 40.90 & 21.45 &  127.35 $\pm$ 0.05 \\
\hline
~500     & 7209.15 & 55.75 & 25.05 &  127.03 $\pm$ 0.05 \\
\hline
~800     & 7409.01 & 69.42 & 28.37 &  126.73  $\pm$ 0.05 \\
\hline
1000     & 7503.90 & 75.91 & 29.94 &  126.59  $\pm$ 0.05 \\
\hline
5000     & 8188.22 &122.72 & 41.24 &  125.57  $\pm$ 0.05 \\
\hline
\end{tabular}
\caption{Numerical results for $\Delta\alphah(\mu^2)$, expressed
  in units $10^{-5}$, for different values of the energy scale $\mu$.
(same input parametes as for table \ref{numbersmz}).
}\label{numbershigher}
\end{table}
The relevant results for an $e^+e^-$ linear collider or at the LHC are
presented in table \ref{numbershigher} .
We observe again that the full 2-loop EW corrections are not
negligible, if compared to the present error on $\Delta\alpha_{had}^{(5)}$.
\\
{\bf The limits on the Higgs boson mass}\\
Since the Higgs boson has not yet been observed, all theoretical
predictions depend parametrically on the value of its mass $m_{\small H}$.
The results of the global fit known as ``blue-band
plot'' prepared by the EW Working Group \cite{EWWG}, 
show the sensitivity of the
$\chi^2$ distribution for the Higgs boson mass to the central value
and to the error as well of $\alpha(\mz)$.
\begin{figure}[h]
\epsfig{file=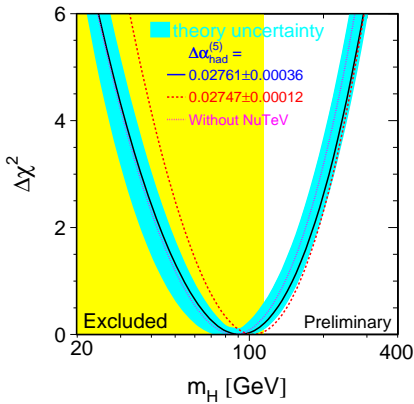,height=5cm,width=5cm,
        bbllx=00pt,bblly=0pt,bburx=500pt,bbury=500pt}
\caption{$\chi^2$ distribution for the Higgs boson mass from the
  global fit to all EW observables (blue-band plot) \cite{EWWG}.}
\label{figbb}
\end{figure}
In fig.\ref{figbb} 
the red and blue lines show the results obtained with two different
  setups for $\Delta\alpha_{had}$; in particular the
  central values differ by $14\cdot 10^{-5}$ parts.
The combined effect of a different choice of the central value and of
a reduction of the error, causes a shift of the minimum of the
distribution by ${\cal O}(15)$ GeV. \\
A detailed study of the limits on $m_{\small H}$ 
has been presented in \cite{dgps}, where
it has been shown that the most sensitive observable to the value of
$m_{\small H}$ is the effective sinus of the weak mixing angle
$\sin^2\theta_{eff}^{lept}$ extracted from the leptonic asymmetries,
whereas the mass of the $W$ boson
has in this respect a weaker constraining power.
On the other hand, the effective sinus is also very sensitive to the
precise value of $\alpha(\mz)$ and to its error as well.\\
At present the determination of the $W$ boson mass is known in the EW
SM at the two-loop level completely \cite{deltar}, whereas only few classes of
two-loop contributions are known for the effective sinus.
A definite answer about the size of the missing corrections to the
effective sinus and about their effect on the limits on the Higgs
boson mass can come only from an explicit two-loop calculation.\\
A global analysis that uses the $\ms$ definition of the weak mixing angle, 
could consistently use the $\ms$ effective coupling $\ehs(\mz)$, to
improve the present evaluation.
One could include the effect of the new corrections:
in particular of the ones due to a loop of light-fermions
exchanging a $W$ or $Z$ bosons, of the so called sub-leading corrections
in the expansion of the top contributions in powers of $m_t$
and of the bosonic diagrams.
Their effect on the indirect limit on the Higgs
boson mass can be roughly estimated to be a reduction of ${\cal O}(6-8)$ GeV
for the 95\% C.L.\ .
Obviously, this is just an estimate which can not replace, by any
means, the result of the global fit to all EW observables.
On the other hand, in the $\ms$ scheme we can consistently compare the
size of the error on $\Delta\alpha_{had}$ to that of the 2-loop weak
corrections and conclude that they are not completely negligible.\\
A similar statement, either positive or negative,
can not be drawn in the on-shell scheme,
because of the gauge-invariance problem which forbids the use of a
resummed effective coupling including the weak effects. In the
on-shell scheme only a complete two-loop calculation can provide a
clean answer about the size and the effect of the missing two-loop corrections.

\section{Conclusions}
We have described the main features of the calculation of the complete
2-loop EW corrections in the SM to the renormalization of the electric
charge. We discussed the computational advantages of the BFM, which
makes evident that the charge renormalization is due only
to the effect of the photon vacuum polarization, whose bosonic
contributions can be Dyson resummed in a manifestly gauge invariant way.
The $\ms$ effective coupling $\ehs(\mz)$, evaluated at the scale $q^2=\mz$,
is shifted by the full 2-loop EW radiative corrections by $+0.00013$,
where 5.8 parts are of purely EW origin.
The electric charge renormalization receives, at the 2-loop level,
QCD, QED and purely EW contributions, which are comparable in size and
with the same sign.\\
In view of an improvement of the measurement of the low-energy
cross-section for $e^+e^- \to~hadrons$, which would imply a reduction
of the error of $\Delta\alpha_{had}$, we observe that the effect of
the perturbative 2-loop EW corrections to the effective electric
charge can not be neglected in all precision tests of the SM.
In particular the indirect limits on the Higgs boson mass are very
sensitive to the precise value of $\alpha(\mz)$.\\
{\bf Acknowledgments}\\
I whish to thank the organizers for the kind invitation and
for the very warm and stimulating atmosphere of the workshop.
I thank Giuseppe Degrassi for his precious and careful comments during
our collaboration and in the preparation of this manuscript.


\begin{thebibliography}{99}

\bibitem{baur}
U. Baur et al., [arXiv:hep-ph/0111314]

\bibitem{pdg}
K.~Hagiwara {\it et al.}  [Particle Data Group Collaboration],
Phys.\ Rev.\ D {\bf 66} (2002) 010001.

\bibitem{procFJ}
see F. Jegerlehner, these proceedings

\bibitem{dv}
G. Degrassi and A. Vicini, 
``Two-loop electric charge renormalization in the Standard Model''
[arXiv:hep-ph/0307122]


\bibitem{sir80}
A.~Sirlin,
Phys.\ Rev.\ D {\bf 22} (1980) 971.


\bibitem{marsir}
W.~J.~Marciano and A.~Sirlin,
Phys.\ Rev.\ D {\bf 22} (1980) 2695
[Erratum-ibid.\ D {\bf 31} (1985) 213].

\bibitem{deltar}
A.~Freitas, W.~Hollik, W.~Walter and G.~Weiglein,
Phys.\ Lett.\ B {\bf 495} (2000) 338
[Erratum-ibid.\ B {\bf 570} (2003) 260]
[arXiv:hep-ph/0007091].

A.~Freitas, W.~Hollik, W.~Walter and G.~Weiglein,
Nucl.\ Phys.\ B {\bf 632} (2002) 189
[Erratum-ibid.\ B {\bf 666} (2003) 305]
[arXiv:hep-ph/0202131].

M.~Awramik and M.~Czakon,
Phys.\ Rev.\ Lett.\  {\bf 89}, 241801 (2002)
[arXiv:hep-ph/0208113].

A.~Onishchenko and O.~Veretin,
Phys.\ Lett.\ B {\bf 551} (2003) 111
[arXiv:hep-ph/0209010].

M.~Awramik, M.~Czakon, A.~Onishchenko and O.~Veretin,
Phys.\ Rev.\ D {\bf 68} (2003) 053004
[arXiv:hep-ph/0209084].

M.~Awramik and M.~Czakon,
Phys.\ Lett.\ B {\bf 568} (2003) 48
[arXiv:hep-ph/0305248].

M.~Awramik, M.~Czakon, A.~Freitas and G.~Weiglein,
arXiv:hep-ph/0311148.

\bibitem{BFM}
B.~S.~Dewitt,
Phys.\ Rev.\  {\bf 162}, 1195 (1967);

J.~Honerkamp,
Nucl.\ Phys.\ B {\bf 48}, 269 (1972);

H.~Kluberg-Stern and J.~B.~Zuber,
Phys.\ Rev.\ D {\bf 12}, 482 (1975);




L.~F.~Abbott,
Nucl.\ Phys.\ B {\bf 185} (1981) 189.

\bibitem{ddw}
A.~Denner, G.~Weiglein and S.~Dittmaier,
Nucl.\ Phys.\ B {\bf 440} (1995) 95
[arXiv:hep-ph/9410338].
A.~Denner and S.~Dittmaier,
Phys. Rev. D {\bf 54} (1996) 4499


\bibitem{allworld}
S.~Eidelman and F.~Jegerlehner,
Z.\ Phys.\ C {\bf 67}, 585 (1995)
[arXiv:hep-ph/9502298];

H.~Burkhardt and B.~Pietrzyk,
Phys.\ Lett.\ B {\bf 356}, 398 (1995), 
Phys.\ Lett.\ B {\bf 513}, 46 (2001);


S.~Groote, J.~G.~Korner, K.~Schilcher and N.~F.~Nasrallah,
Phys.\ Lett.\ B {\bf 440}, 375 (1998)
[arXiv:hep-ph/9802374];

M.~Davier and A.~H\"ocker,
Phys.\ Lett.\ B {\bf 419}, 419 (1998)
[arXiv:hep-ph/9711308];

J.~H.~K\"uhn and M.~Steinhauser,
Phys.\ Lett.\ B {\bf 437}, 425 (1998)
[arXiv:hep-ph/9802241];

A.~D.~Martin and D.~Zeppenfeld,
Phys.\ Lett.\ B {\bf 345}, 558 (1995)
[arXiv:hep-ph/9411377];

A.~D.~Martin, J.~Outhwaite and M.~G.~Ryskin,
Phys.\ Lett.\ B {\bf 492}, 69 (2000)
[arXiv:hep-ph/0008078].


\bibitem{fks}
S.~Fanchiotti, B.~A.~Kniehl and A.~Sirlin,
Phys.\ Rev.\ D {\bf 48} (1993) 307
[arXiv:hep-ph/9212285].

\bibitem{dg}
G.~Degrassi and P.~Gambino,
Nucl.\ Phys.\ B {\bf 567} (2000) 3
[arXiv:hep-ph/9905472].


\bibitem{newexp}
see Eidelman's, Hu's, Dytman's and\\ 
Valeriani's talks,  these proceedings and\\
http://www.pi.infn.it/congressi/sighad03/

\bibitem{jegerlehner}
F.~Jegerlehner, hep-ph/0310234

\bibitem{dh}
M.~Davier and A.~Hocker,
Phys.\ Lett.\ B {\bf 435} (1998) 427
[arXiv:hep-ph/9805470].


\bibitem{dgv}
G.~Degrassi, S.~Fanchiotti and P.~Gambino,
Int.\ J.\ Mod.\ Phys.\ A {\bf 10}, 1377 (1995)
[arXiv:hep-ph/9403250];

G.~Degrassi, S.~Fanchiotti, F.~Feruglio, P.~Gambino and A.~Vicini,
Phys.\ Lett.\ B {\bf 350}, 75 (1995)
[arXiv:hep-ph/9412380];

G.~Degrassi, P.~Gambino and A.~Vicini,
Phys.\ Lett.\ B {\bf 383} (1996) 219
[arXiv:hep-ph/9603374].



\bibitem{dgps}
A. Ferroglia, G. Ossola, M. Passera and A. Sirlin, Phys. Rev. D
{\bf 65} (2002) 113002

\bibitem{EWWG}
LEP Electroweak Working Group, http://lepewwg.web.cern.ch/LEPEWWG/




\end{thebibliography}
\end{document}